\documentclass{elsart}

\usepackage{graphicx}
\journal{Physica A}
\newcommand{\erw} [1] {\ensuremath{\langle #1 \rangle}}
\begin{document}
\begin{frontmatter}

\title{Comparison of detrending methods for fluctuation analysis}
\author[Bar-Ilan]{Amir Bashan},
\author[Bar-Ilan]{Ronny Bartsch\corauthref{cor1}},
\ead{bartsch.ronny@gmail.com}
\corauth[cor1]{Corresponding author. Tel.: + 972 3 5317885; fax: + 972 3 5317884.}
\author[Halle]{Jan W. Kantelhardt}, and
\author[Bar-Ilan]{Shlomo Havlin}
\address[Bar-Ilan]{Minerva Center, Dept. of Physics, Bar-Ilan University, Ramat-Gan 52900,
Israel}
\address[Halle]{Institute of Physics, Martin-Luther-Universit\"at Halle-Wittenberg, 
06099 Halle, Germany}
\small{submitted: 22 August 2007; accepted: 31 March 2007}

\begin{abstract}
We examine several recently suggested methods for the detection of long-range correlations 
in data series based on similar ideas as the well-established Detrended Fluctuation Analysis 
(DFA). In particular, we present a detailed comparison between the regular DFA and two 
recently suggested methods: the Centered Moving Average (CMA) Method and a Modified Detrended 
Fluctuation Analysis (MDFA). We find that CMA is performing equivalently as DFA in long data with 
weak trends and slightly superior to DFA in short data with weak trends. When comparing
standard DFA to MDFA we observe that DFA performs slightly better in almost all examples 
we studied. We also discuss how several types of trends affect the different types of DFA. 
For weak trends in the data, the new methods are comparable with DFA in these respects. However, if the functional
form of the trend in data is not a-priori known, DFA remains the method of choice. 
Only a comparison of DFA results, using different detrending polynomials, yields full 
recognition of the trends. A comparison with independent methods is recommended for 
proving long-range correlations.
\end{abstract}

\end{frontmatter}

% main text
\section{Introduction}

Experimental data are often affected by non-stationarities, i.e. varying
mean and standard deviation. These effects 
have to be well distinguished from the intrinsic fluctuations and correlations of the 
system in order to find the correct scaling behaviour. Sometimes we do not know 
the reasons for underlying non-stationarities in collected data and -- even worse -- we do not 
know the type of the underlying non-stationarities.  

In the last decade {\it Detrended Fluctuation Analysis} (DFA), originally 
introduced by Peng et al. \cite{peng94}, has been established as an important 
method to reliably detect long-range (auto-) correlations\footnote{In the following 
we will label long-range as long-term when speaking, more specifically, about time series.} in data effected by 
trends. The method is based on random walk theory.  Its non-detrending 
predecessors are Hurst's rescaled range analysis \cite{hurst51} and fluctuation 
analysis (FA) \cite{peng92}.  DFA was later generalized for higher order detrending 
\cite{bunde00}, multifractal analysis \cite{kantelhardt02}, separate analysis 
of sign and magnitude series \cite{ashkenazy01}, and data with more than one
dimension \cite{gu06}.  Its features have been studied in many articles 
\cite{kantelhardt01,hu01,chen02,chen03,grau06,nagarajan06}.  In addition, 
several comparisons of DFA with other methods for stationary and non-stationary 
time-series analysis have been published, see, e.g., 
\cite{taqqu95,heneghan00,Weron02,Mielniczuk07} and in particular 
\cite{delignieresa06}, where DFA is compared with many other established methods
for short data sets. 

The convenience of DFA has led to a broad range of application in very diverse 
fields where long-range correlations are of interest:  
\begin{itemize} 
\item DNA sequences, %\cite{peng94}, 
\item medical and physiological time series (recordings of heartbeat, breathing, blood 
pressure, blood flow, nerve spike intervals, human gait, glucose levels, gene expression 
data),	
%[Yamamoto07,Esen06,Voss05,Nagarajan06]
\item geophysics time series (recordings of temperature, precipitation, water runoff, 
ozone levels, wind speed, seismic events, vegetational patterns, and climate 
dynamics), %[livina07,Telesca07,kurnaz04,kantz04]
\item astrophysical time series (X-ray light sources and sunspot numbers), 
%[zebende03/05, movahed06]
\item technical time series (internet traffic, highway traffic, and neutronic power 
from a reactor), %[Tadaki06,Espinosa-Paredes]
\item social time series (finance and economy, language characteristics, fatalities in 
the Iraq conflict), %[Kosmidis06,Urrea07]
as well as
\item physics data, e.g., surface roughness, %[Feudel06]
chaotic spectra of atoms, %[Santhanam06]
and photon correlation spectroscopy recordings. %[Stadler03-06]
\end{itemize}
Altogether, there are about 450 papers applying DFA. In most cases positive 
auto-correlations were reported leaving only a few exceptions with anti-correlations, 
see, e.g., \cite{bahar01,ekg form,santhanam06}. 

In the DFA technique -- as in all techniques based on random walk theory -- time series 
are integrated by partial summation which enables also the analysis of data with weak correlations.
In addition the integration reduces the noise level caused by imperfect measurements and
noise, an advantage that applies also to other related non-detrending methods 
\cite{hurst51,peng92}. However, for the reliable characterization of time series, it 
is also essential to distinguish trends from intrinsic fluctuations, that 
might be long-term correlated. Monotonous, periodic or step-like 
trends are caused by external effects, e.~g., by the greenhouse warming 
\cite{greenhouse1}, seasonal 
variations for temperature records \cite{ekb98} and river runoffs \cite{hurst51,kantelhardt03,ekb06,runoff}, 
different levels of daily activity in long-term 
physiological data \cite{karasik}, or unstable light sources in photon correlation spectroscopy \cite{jan_PRB}.
To characterize a complex system based on time series, trends and fluctuations are usually 
studied separately (see, e.g., \cite{schmitt06} for a recent discussion). Strong trends in 
data can lead to a false detection of long-term correlations if only one (non-detrending) 
method is used or if the results are not carefully interpreted. A major advantage of the 
DFA technique is the systematic elimination of polynomial trends of different order 
\cite{bunde00,kantelhardt01,hu01}. Note however that an additive composition of fluctuations
and trends is assumed. The technique can thus assist in gaining insight 
into the scaling behaviour of the natural variability as well as into the kind of trends 
of the considered time series \cite{greenhouse2}.  

Still, we would like to note that conclusions should not be based on DFA or variants 
of this method alone in most applications. In particular, if it is not clear whether a 
given time series is indeed long-term correlated or just short-term correlated with a 
fairly large correlation time scale, results of DFA should be compared with other methods.  
For example, one can employ wavelet methods (see, e.g., 
\cite{ekb98,ekb06,Audit-02,Ignaccolo-04,Principato-07}). Another option is to remove 
short-term correlations by considering averaged series for comparison.  For a time 
series with daily observations and possible short-term correlations up to two years, 
for example, one might consider the series of two-year averages and apply DFA as well as 
FA, Hurst's Analysis, binned power spectra analysis, and/or wavelet analysis.  Only if 
at least two independent methods consistently indicate long-term correlations, one can 
be sure that the data are indeed long-term correlated.

Lately, several modifications of the DFA method have been suggested with many different 
techniques for the elimination of monotonous and periodic trends.  These methods include 
\begin{itemize}
\item the Detrended Moving Average technique \cite{BMA1,BMA2,BMA3}, which we denote by 
Backward Moving Average (BMA) technique (following \cite{CMA}), 
\item the Centered Moving Average (CMA) method \cite{CMA}, an essentially 
improved version of BMA, 
\item the Modified Detrended Fluctuation Analysis (MDFA) \cite{MDFA}, which is essentially 
a mixture of old FA and DFA, 
\item the continuous DFA (CDFA) technique \cite{CDFA1,CDFA2}, which is particularly intended 
for the detection of transitions, 
\item the Fourier DFA \cite{FourierDFA},
\item a variant of DFA based on empirical mode decomposition (EMD) \cite{EMD-DFA}, 
\item a variant of DFA based on singular value decomposition (SVD) \cite{SVD-DFA1,SVD-DFA2}, 
and
\item a variant of DFA based on high-pass filtering \cite{HP-DFA}.
\end{itemize}
Although several of the original publications compare their new suggested method
with the DFA, there is no inter-comparison between these new methods.  Hence, it is not 
clear which methods might be most suitable for which application.  In this work we 
comment on all recently suggested detrending random walk based methods we are aware of.  
Moreover, we study and compare in detail two of the most prominent and -- according to 
our studies -- most suitable new methods with standard DFA, presenting their advantages 
and disadvantages.  For recent comparative studies not focused on detrending methods, see 
\cite{taqqu95,Mielniczuk07,delignieresa06}.  For studies comparing DFA and BMA, see 
\cite{grech05,xu05}; note that \cite{xu05} also discusses CMA.  For studies comparing 
methods for detrending multifractal analysis (multifractal DFA (MF-DFA) and wavelet 
transform modulus maxima (WTMM) method), see \cite{kantelhardt02,kantelhardt03,oswiecimka06}.

The paper is organized as follows: In Section 2 we thoroughly explain the standard DFA 
as well as the Centered Moving Average (CMA) method, and a Modified Detrended Fluctuation 
Analysis (MDFA). We further introduce several other (more complicated) detrending methods
and remark on their utility.  Section 3 reports and discusses our results for DFA, CMA and 
MDFA, obtained from monofractal artificial time series with different lengths, crossovers 
and monotonous trends. We conclude in Section 4.

\section{Methods}

\subsection{Long-Range Correlations}

We consider a record $(x_i)$ of $i=1,\ldots,N$ equidistant measurements.  In 
most applications, the index $i$ will correspond to the time of the measurements.
We are interested in the correlation of the values $x_i$ and $x_{i+s}$ 
for different time lags, i.~e. correlations over different time scales $s$.  
In order to remove a constant offset in the data, the mean 
$\langle x \rangle = {1 \over N} \sum_{i=1}^{N} x_i$ is usually subtracted, 
$\tilde {x}_i \equiv x_i - \langle x \rangle$.  Quantitatively, correlations 
between $x$-values separated by $s$ steps are defined by the (auto-) correlation 
function
\begin{equation} C(s) = \frac{\big\langle \tilde{x}_i \, \tilde{x}_{i+s}\big\rangle}
{\big\langle \tilde{x}_i^2\big\rangle} 
= {1 \over (N-s)\big\langle \tilde{x}_i^2\big\rangle} \sum_{i=1}^{N-s} \tilde{x}_i \, \tilde{x}_{i+s}.
\label{autocorr}\end{equation}
If the $x_i$ are uncorrelated, $C(s)$ is zero for $s>0$.  Short-range correlations 
of the $x_i$ are described by $C(s)$ declining exponentially, $C(s) \sim \exp(-s/
t_\times)$ with a decay time $t_\times$.  For so-called {\it long-range correlations}, 
$t_\times=\int_0^\infty C(s)\, {\rm d}s$ diverges and the decay time $t_\times$ cannot 
be defined. For example, $C(s)$ declines as a power-law
\begin{equation} C(s) \sim s^{- \gamma} 
\label{gamma}
\end{equation}
with an exponent $0 < \gamma < 1$. A direct calculation of $C(s)$ is usually not 
appropriate due to 
underlying non-stationarities and trends of unknown origin.  
%This makes 
%the definition of $C(s)$ problematic, because the average $\langle x \rangle$ is not
%well-defined.  
Furthermore, $C(s)$ strongly fluctuates around zero on large scales $s$, making it
impossible to find the potential scaling behaviour (\ref{gamma}). 
Thus, one has to determine the correlation exponent $\gamma$ indirectly.

Note that in some applications a separate inspection of short-term and long-term 
correlations is desirable. A convenient way to exclude short-term correlations
up to a scale $s_l$ is downsampling the original data by the same factor $s_l$. Contrariwise, the segmentation
of the data into boxes of length $s_u$ and a subsequent shuffling of the boxes destroys long-term
correlations on scales above $s_u$.

\subsection{Detrended Fluctuation Analysis (DFA)}

The method of {\it Detrended Fluctuation Analysis} (DFA) \cite{peng94} is an 
improvement of classical fluctuation analysis (FA) \cite{peng92}, which is similar 
to Hurst's rescaled range ($R/S$-) analysis \cite{hurst51}. They allow determining 
the correlation properties on large time scales.  All three methods are based on 
random walk theory.  One first calculates the `profile' 
\begin{equation} X(n)=\sum_{i=1}^n (x_i-\erw{x}) \end{equation}
of a time series $(x_i)$, $i=1,\ldots,N$ (with mean $\erw{x}$), which can be considered as 
the position of a random walker on a linear chain after $n$ steps. 

Then the profile is divided into $N_s\equiv [N/s]$ non-overlapping segments of equal length 
(`scale') $s$. The mean-squared fluctuation function of the FA method is given by 
\begin{equation} F^2(s)={1 \over N_s}\sum_{\nu=1}^{N_s} [X((\nu-1)s) - X(\nu s)]^2. 
\end{equation}
In Hurst's $R/S$ analysis (see \cite{Mielniczuk07} and references therein for a recently 
suggested improved version and tests), one calculates in each segment $\nu$ the range 
$R$ of $X(n)$ given by the difference between maximal and minimal value, 
$R(s)=X_{\rm max}-X_{\rm min}$. 
The ``rescaling of range'' is done by dividing $R(s)$ by the corresponding standard deviation $S(s)=\sigma(X(n))$ 
of the same segment $\nu$. The mean of all quotients at a particular scale $s$ is equivalent 
to $F(s)$ (except for multi-fractal data) and usually shows a power-law scaling relationship 
with $s$.

While both, FA and Hurst's method fail to determine correlation properties if linear or 
higher order trends are present in the data, %or the original time series is non-stationary, 
DFA explicitly deals with monotonous trends in a detrending procedure. This is done by 
estimating a piecewise polynomial trend $y_{s}^{(p)}(n)$ within each segment $\nu$ by 
least-square fitting. I.e., $y_{s}^{(p)}(n)$ consists of concatenated polynomials of order 
$p$ which are calculated separately for each of the segments. The detrended profile function $\tilde{X}_s(n)$ 
on scale $s$ is determined by (`detrending'):
\begin{equation} \tilde{X}_s(n)=X(n)-y_s^{(p)}(n). \label{detrendedfunction} \end{equation}
The degree of the polynomial can be varied in order to eliminate linear $(p=1)$, quadratic 
$(p=2)$ or higher order trends of the profile function \cite{bunde00}. Conventionally 
the DFA is named after the order of the fitting polynomial (DFA1, DFA2, ...).
Note that DFA1 is equivalent to Hurst's analysis in terms of detrending.

The variance of $\tilde{X}_s(n)$ yields the fluctuation function on scale $s$
\begin{equation} F(s)=\left\{\frac{1}{N}\sum_{n=1}^{N} \tilde{X}_s^2(n) \right\}^{1/2}.
\label{fluctuationfunction} \end{equation}
This function, which has to be calculated for different scales $s$, corresponds to the 
trend-eliminated root mean square displacement of the random walker mentioned above and is 
related to the auto-correlation function by an integral expression \cite{peng94}; see the 
appendix of \cite{taqqu95} for a derivation for DFA1. For an equivalent, but more common 
description of DFA, see, e.g., \cite{kantelhardt01}. We note that in 
studies that include averaging over many records (or one record cut into many separate 
pieces by the elimination of some unreliable intermediate data points) the averaging 
procedure (\ref{fluctuationfunction}) must be performed for all data. Taking the square
root should usually be the final step after all averaging is finished; however note
\cite{Weron02,Mielniczuk07}, where this order is reversed. It is usually not appropriate 
to calculate $F(s)$ for parts of the data and then average the $F(s)$ values, since such 
a procedure will bias the results towards smaller scaling exponents on large time scales
close to the maximum scale $s_{\rm max} \approx N/4$ where statistically reliable results 
can be obtained \cite{kantelhardt01}.

If $F(s)$ increases for increasing $s$ asymptotically as 
\begin{equation} F(s)\sim s^\alpha \end{equation}
with $0.5<\alpha<1$, one finds that the scaling (or 'Hurst') exponent $\alpha$ is related 
to the correlation exponent $\gamma$ by $\alpha=1-\gamma/2$ \cite{taqqu95}. A value of 
$\alpha=0.5$ thus indicates 
that there are no (or only short-term) correlations. If $\alpha>0.5$, the data are 
long-term correlated. The higher $\alpha$, the stronger are the correlations in the signal.
Note that $\alpha>1$ indicates a non-stationary local average of the data; in this 
case both, FA and Hurst analysis fail and yield only $\alpha=1$. The case $\alpha<0.5$
corresponds to long-term anticorrelations, meaning that large values are most likely to be 
followed by small values and vice versa \cite{bahar01,ekg form,santhanam06}.

If the type of trends in given data is not known beforehand, the fluctuation function $F(s)$
should be calculated for several orders $p$ of the fitting polynomial. If $p$ is too 
low, $F(s)$ will show a pronounced crossover to a regime with larger slope for large scales 
$s$ \cite{kantelhardt01,hu01}. The maximum slope of $\log F(s)$ versus $\log s$ is $p+1$.  
The crossover will move to larger scales $s$ or disappear when $p$ is increased, unless it 
is a real crossover in the intrinsic fluctuations and not due to trends \cite{kantelhardt01}. 
Hence, one can find $p$ such that detrending is 
sufficient. However, $p$ should not be larger than necessary, because deviations on
short scales $s$ increase with increasing $p$.

\subsection{Centered Moving Average (CMA) Analysis}

A possible drawback of the DFA method is the occurrence of abrupt jumps in the detrended profile $\tilde{X}_s(n)$ 
(Eq.~(\ref{detrendedfunction})) at the boundaries between the segments, since the fitting 
polynomials in neighbouring segments are not related. A simple way to avoid these jumps would 
be the calculation of $F(s)$ based on polynomial fits in overlapping windows. However, this 
is rather time consuming due to the polynomial fit in each segment and is consequently not 
done in most applications. To overcome the problem of artificial jumps and to reliable 
determine the scaling exponent $\alpha$ in non-stationary time series, several 
modifications of the FA and DFA methods were suggested in the last years. 

A particular attractive modification leads to the methods of {\it Detrended Moving Average} 
(DMA), where running averages replace the polynomial fits. Its first suggested version, the 
{\it Backward Moving Average} (BMA) method \cite{BMA1,BMA2,BMA3}, however, slightly 
underestimates the scaling exponent by about 0.05, because an artificial time shift of 
$s$ between the original signal and the moving average is introduced.  This time shift 
leads to an additional contribution to $\tilde{X}_s(n)$, which causes a larger fluctuation 
function $F(s)$ in particular for small scales in the case of long-term correlated data \cite{grech05}.  
In addition, the BMA method is effectively not detrending \cite{xu05}.  
Its slope $\alpha$ is limited by 1 just as for the earlier non-detrending methods FA and $R/S$.

It was soon recognized that the artificial time shift of the BMA method can easily be eliminated. 
This leads to the {\it Centered Moving Average} 
(CMA) method \cite{CMA}, where $\tilde{X}_s(n)$ is calculated as 
\begin{equation} \tilde{X}_s(n)=X(n)-\frac{1}{s}\sum_{j=-(s-1)/2}^{(s-1)/2}X(n+j),
\label{CMA} \end{equation}
while Eq.~(\ref{fluctuationfunction}) stays the same. Unlike DFA, the CMA method cannot 
easily be generalized to remove linear and higher order trends in the data.
However, CMA is somehow similar to DFA1 with overlapping windows.    

\subsection{Modified Detrended Fluctuation Analysis (MDFA)}

Another type of detrended fluctuation analysis, which we will denote as {\it Modified 
Detrended Fluctuation Analysis} (MDFA) \cite{MDFA}, eliminates trends similar to the 
DFA method. A polynomial is fitted to the profile function $X(n)$ in each segment $\nu$ 
and the deviation between the profile function and the polynomial fit is calculated, 
$\tilde{X}_s(n)=X(n)-y_s^{(p)}(n)$.  To estimate correlations in the data, this method 
uses a derivative of $\tilde{X}_s(n)$, obtained for each segment $\nu$, by 
$\Delta\tilde{X}_s(n)=\tilde{X}_s(n+s/2)-\tilde{X}_s(n)$. Hence, 
Eq.~(\ref{fluctuationfunction}) becomes 
\begin{equation} F(s)=\left\{\frac{1}{N}\sum_{n=1}^{N} [\Delta\tilde{X}_s(n)]^2 \right\}^{1/2}.
\label{fluctuationfunctionJFA} \end{equation}  
As in case of DFA, MDFA can easily be generalized to remove higher order trends in the data.
Since the fitting polynomials in adjacent segments are not related, $\Delta\tilde{X}_s(n)$
shows abrupt jumps on their boundaries as well. This leads to fluctuations of $F(s)$ for large 
segment sizes $s$ as we will show below.

\subsection{Further Modifications and Extensions of DFA}

Several modifications and extensions of DFA have been proposed.  Most of them are, 
however, rather complicated in implementation. While they might be very useful in particular 
applications, we believe the implications of the complicated detrending and decomposition 
techniques are not sufficiently understood and their programming effort is too large for 
a wide usage.

The {\it Fourier-detrended fluctuation analysis} \cite{FourierDFA} aims to eliminate 
slow oscillatory trends which are found especially in weather and climate series due to 
seasonal influences. The character of these trends can be rather periodic and regular or 
irregular, and their influence on the detection of long-range correlations by means of 
DFA was systematically studied previously \cite{kantelhardt01}. Among other things it has 
been shown that low-frequency periodic trends disturb the scaling behaviour of the results 
much stronger than high-frequency trends and thus have to be removed prior to the 
analysis. In case of periodic and regular oscillations, e.g., in temperature fluctuations 
one simply removes the low frequency seasonal trend by subtracting the daily mean 
temperatures from the data. Another way, which the Fourier-detrended fluctuation 
analysis suggests, is to filter out the relevant frequencies in the signals' Fourier 
spectrum before applying DFA to the filtered signal. Nevertheless, this method which is only an extension of DFA faces 
several difficulties especially its limitation to periodic and regular trends.
Furthermore one needs to know the interfering frequency band beforehand. 

To study correlations in data with quasi-periodic or irregular oscillating trends, {\it 
empirical mode decomposition} (EMD) was suggested \cite{EMD-DFA}. 
The EMD algorithm breaks down the signal into its intrinsic mode functions (IMFs) which 
can be used to distinguish between fluctuations and trends. The trends, estimated 
by a quasi-periodic fit containing the dominating frequencies of a sufficiently large 
number of IMFs, is subtracted from the data, yielding a slightly better scaling behaviour 
in the DFA curves. However, we believe, that this extension of DFA is too complicated for wide-spread
applications.

Another extension of DFA which was shown to minimize the effect of periodic and quasi-periodic 
trends is based on {\it singular value decomposition} (SVD) \cite{SVD-DFA1, SVD-DFA2}. In 
this approach, one first embeds the original signal in a matrix whose dimension has to be 
much larger than the number of frequency components of the periodic or quasi-periodic 
trends obtained in the power spectrum. Applying SVD yields a diagonal matrix which can be 
manipulated by setting the dominant eigen-values (associated with the trends) to zero. 
The filtered matrix finally leads to the filtered data, and it has been shown that subsequent 
application of DFA determines the expected scaling behaviour if the embedding dimension
is sufficiently large. None the less, the performance of this rather complex method seems 
to decrease for larger values of the scaling exponent. Furthermore SVD-DFA assumes that trends 
are deterministic and narrow banded.

Nevertheless the above-mentioned extensions of DFA show the need for a fluctuation analysis 
that can also handle oscillatory data automatically. The detrending procedure 
in DFA (Eq.~(\ref{detrendedfunction})) can be regarded as a 
scale-dependent high-pass filter since (low-frequency) fluctuations exceeding a specific 
scale $s$ are eliminated. Therefore, it has been suggested to obtain the detrended profile 
$\tilde{X}_s(i)$ for each scale $s$ directly by applying digital high-pass filters 
\cite{HP-DFA}. In particular, Butterworth, Chebyshev-I, Chebyshev-II, and an elliptical 
filter were suggested. While the elliptical filter showed the best performance in detecting 
long-range correlations in artificial data, the Chebyshev-II filter was found to be 
problematic. Additionally, in order to avoid a time shift between filtered and original 
profile, the average of the directly filtered signal and the time reversed filtered signal 
is considered. The effects of these complicated filters on the scaling behaviour are, 
however, not fully understood.% We thus do not recommend the method for data with unknown
%properties.

Finally, a continuous DFA method has been suggested in the context of studying heartbeat
data during sleep \cite{CDFA1,CDFA2}.  The method compares unnormalized fluctuation functions 
$F(s)$ for increasing length of the data.  I.e., one starts with a very short recording and
subsequently adds more points of data.  The method is particularly suitable for the detection
of change points in the data, e.g., physiological transitions between different activity or
sleep stages.  Since the main objective of the method is not the study of scaling behaviour,
we do not discuss it in detail in this comparison.

\section{Results}

\subsection{Estimating the scaling behaviour in long and short data sets}

\begin{figure} [!ht] \begin{center}
\includegraphics[height=5.05cm]{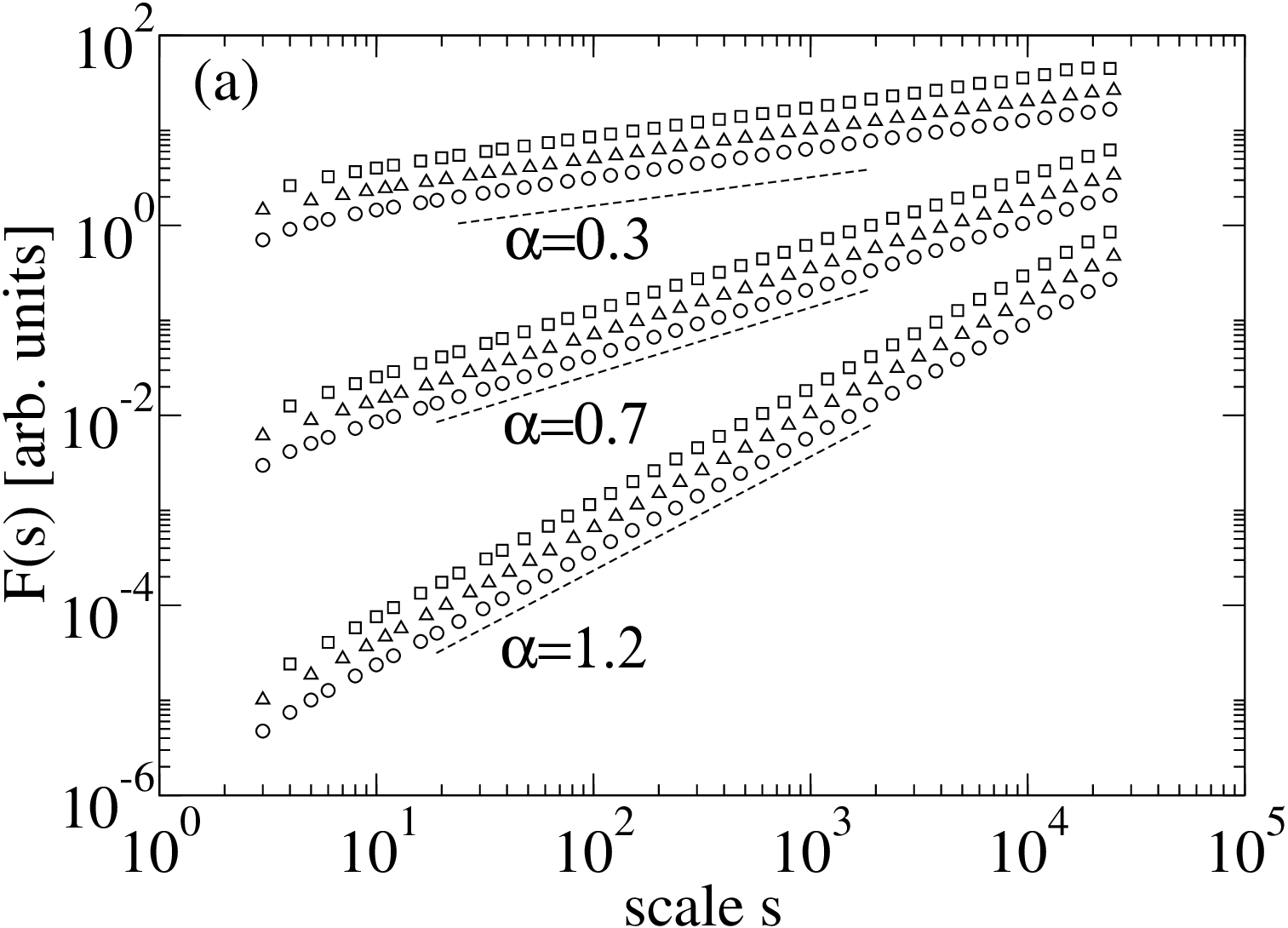}\hfill\includegraphics[height=4.9cm]{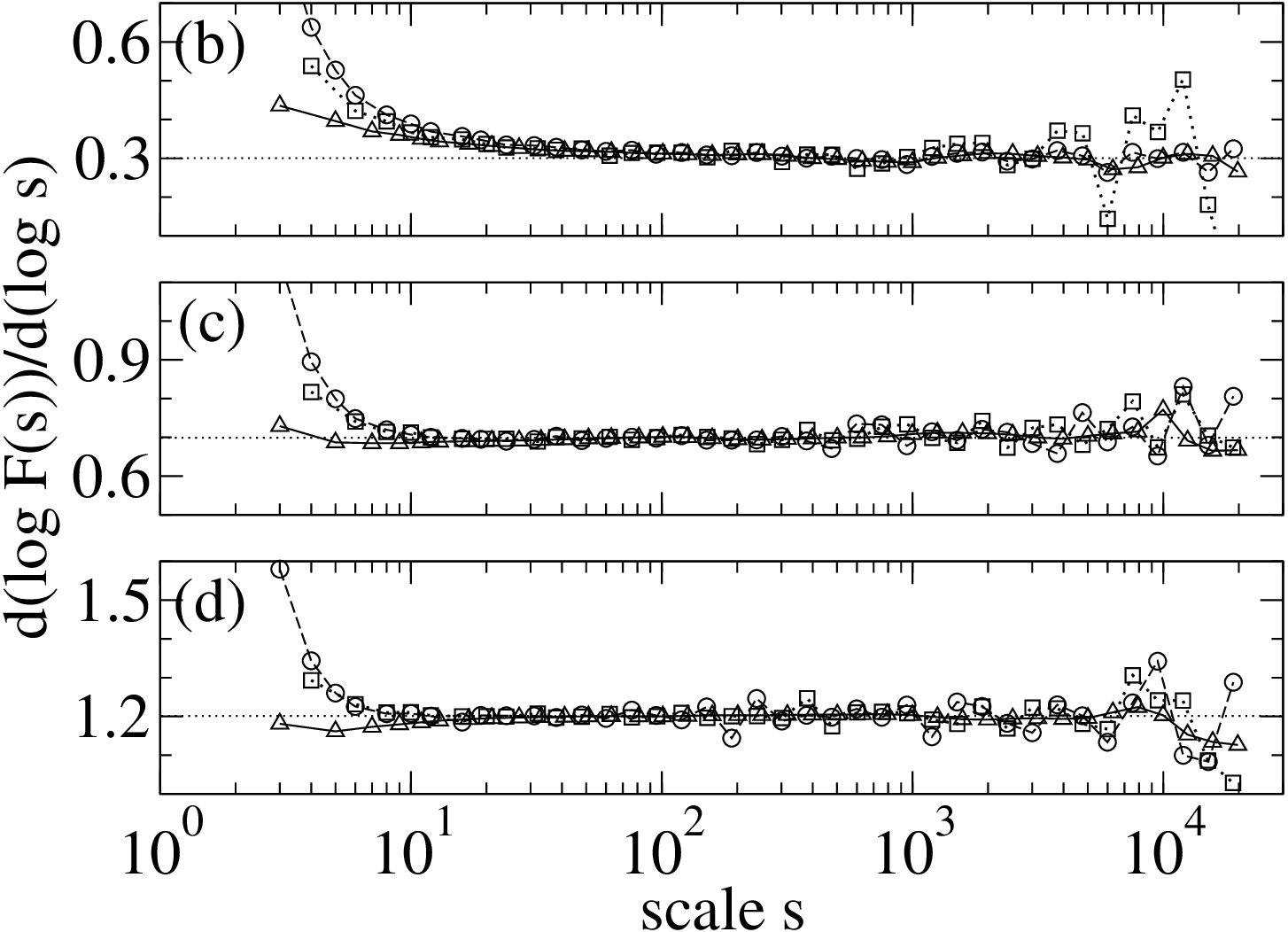}
\caption{(a) Fluctuation functions $F(s)$ versus scale $s$ of long-range correlated 
data with different scaling exponents $\alpha=0.3, 0.7, 1.2$, by means of DFA1 (circles), CMA
(triangles) and MDFA1 (squares). The results have been obtained by 
averaging $F^2(s)$ over 100 artificial series of length $N=50000$ 
for each method and scaling exponent. The DFA1 curve is shifted downwards for clarity. 
(b)-(d) Point to point derivative of the average fluctuation functions
shown in Fig. \ref{long}(a) for (b) $\alpha=0.3$, (c) $\alpha=0.7$ and (d) $\alpha=1.2$. 
Note the deviation from the scaling behaviour for small and large scales for DFA1 and MDFA1.}
\label{long} \end{center} \end{figure}

In the first part of our comparison between DFA, CMA and MDFA, we calculate the scaling 
exponent $\alpha$ for long-range correlated normally distributed data sets of length $N=50000$. The data sets 
are generated using the modified Fourier filtering method, see, e.~g., \cite{makse}. As 
one can see in Fig. \ref{long}(a), all three methods give sufficiently good results for 
different values of $\alpha$. However, on closer examination, i.~e., looking at the 
successive slopes (logarithmic point to point derivatives) of $F(s)$ 
(Figs. \ref{long}(b)-(d)), it can be seen that DFA1 and MDFA1 systematically overestimate 
the scaling exponent for small scales $s$. This effect has already been discussed for 
DFA and a modification was suggested for removing this artifact 
\cite{kantelhardt01,parkinson}. In addition, the significant fluctuations of the 
successive slopes of DFA1 (and MDFA1) on large scales $s$, led to the rule of determining 
$\alpha$ only up to a scale of $N/4$ \cite{kantelhardt01}. Nevertheless, Figs.
\ref{long}(b)-(d) show that the scaling behaviour of CMA is more stable than for DFA1 
and MDFA1, suggesting that CMA could be used for reliable computation of $\alpha$ even 
for scales $s<10$ and up to $s_{\rm max} = N/2$. 

\begin{figure} [ht] \begin{center}
\includegraphics[height=9cm]{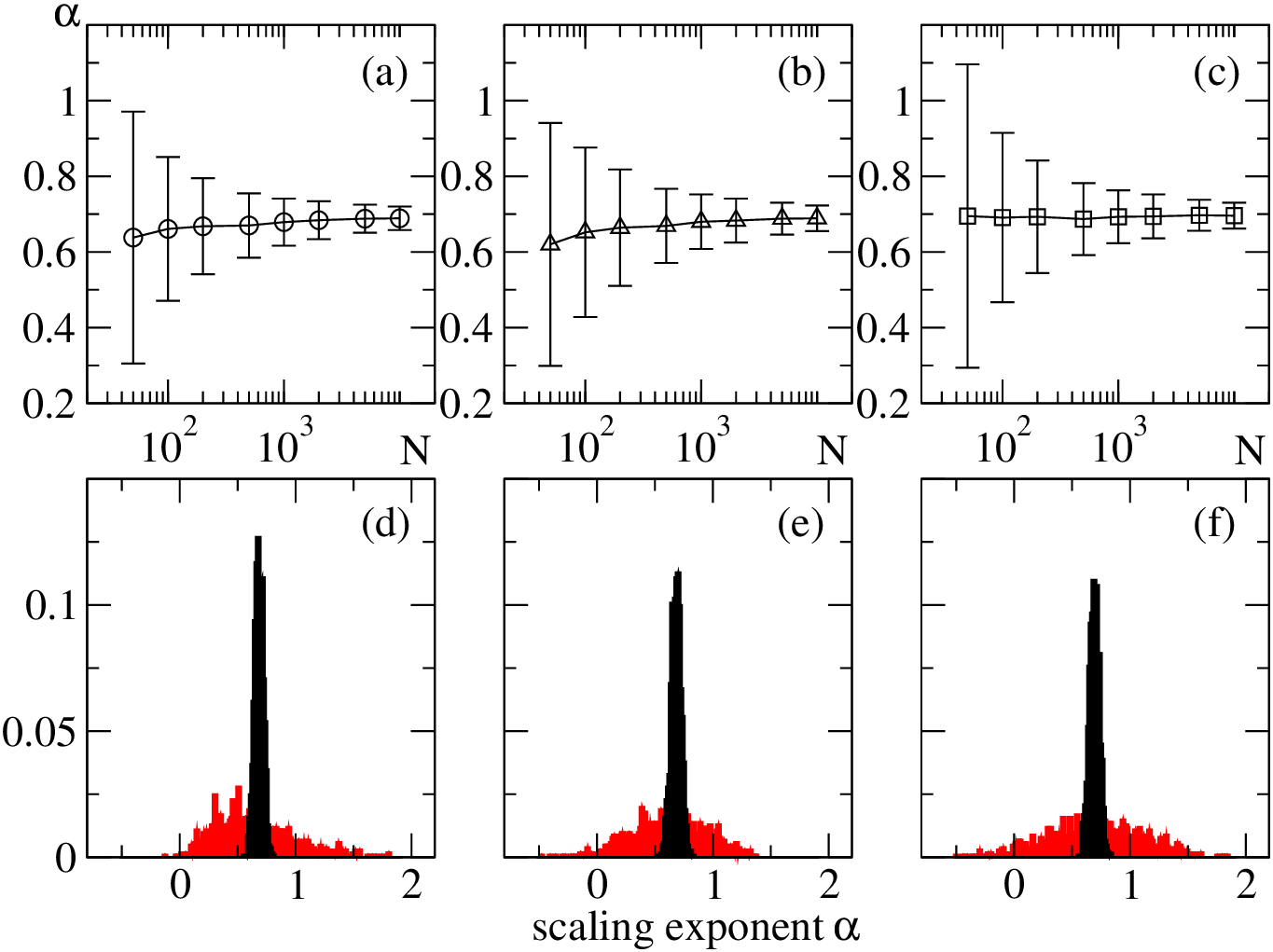}
\caption{(colour online) Mean and standard deviation of the calculated scaling exponents for 
different signal lengths $N$ and histograms of scaling exponents. 
The fitting range was fixed at a lower limit namely $s\in [10,N/2]$.
We applied (a) DFA1, (b) CMA and (c) MDFA1 to 1000 generated series with $\alpha=0.7$. 
In (d)-(f), the corresponding normalized histograms of the scaling exponents 
are shown for $N=50$ (red colour) and $N=5000$ (black colour). Note that the distribution 
of $\alpha_{\rm DFA}$ is asymmetric for $N=50$.}
\label{short} \end{center} \end{figure}

\begin{figure} [ht] \begin{center}
\includegraphics[height=6cm]{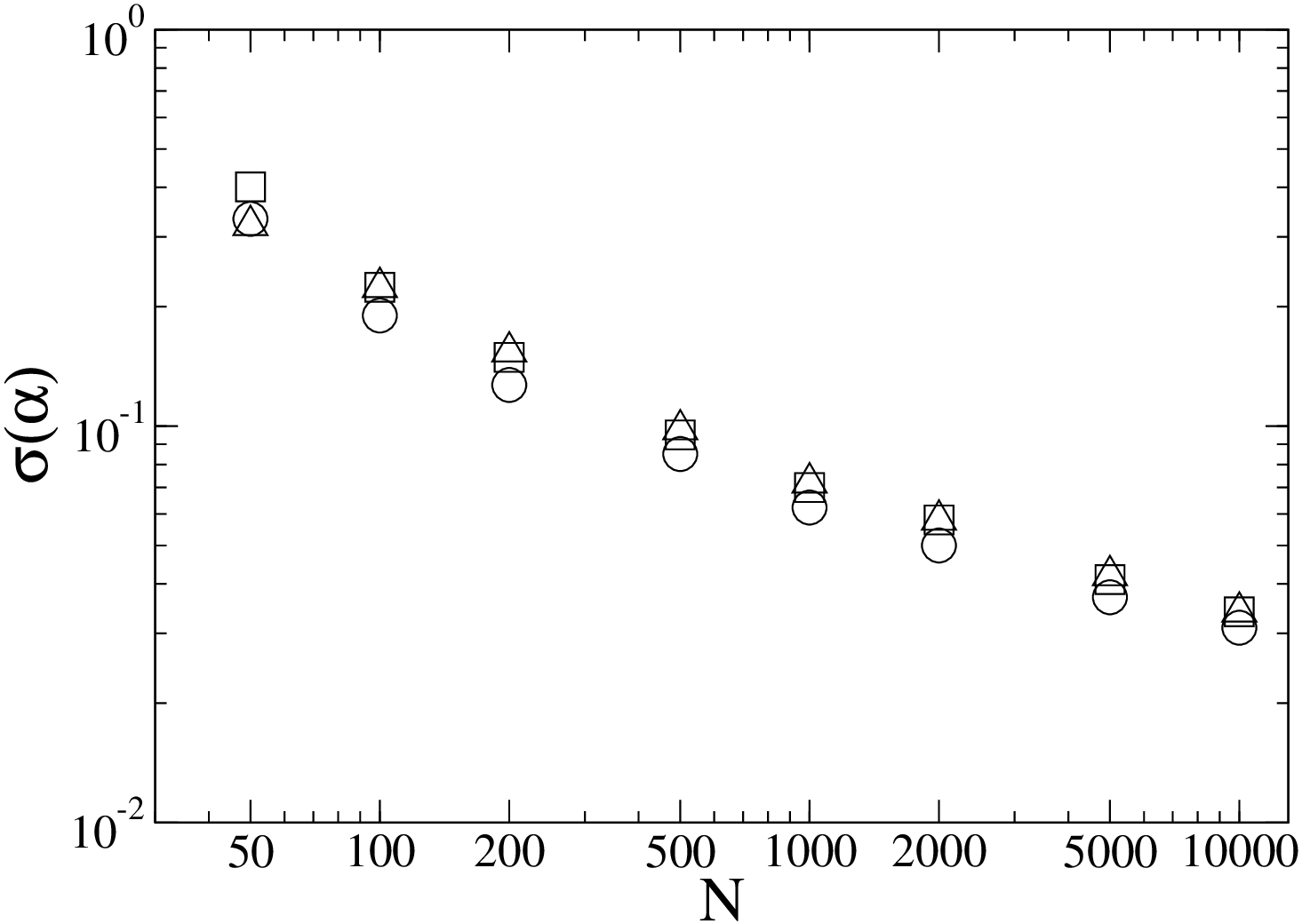}
\caption{Standard deviation of $\alpha$ versus signal lengths $N$ for DFA1 (circles), 
CMA (triangles) and MDFA1 (squares). This figure can be used as a calibration curve, i.e.
to estimate the uncertainty, $\sigma$, of $\alpha$ depending on the signal length $N$. 
For each method we averaged over 1000 series with $\alpha=0.7$.}
\label{SD single} \end{center} \end{figure}

An important topic in fluctuation analysis is the influence of the signal length upon the 
reliability of the estimated scaling behaviour. For this purpose, we applied DFA1, CMA and 
MDFA1 on long-range correlated data and calculated the mean and standard deviation of the scaling
exponents ($\bar{\alpha}$, $\sigma(\alpha)$) as function of the signal length $N$ 
(Figs. \ref{short} (a)-(c)). There are two ways in defining the scaling range for 
the fitting procedure of $\alpha$.  Firstly one can fix the lower limit to $s=10$ (in order
to reduce the influence of the small scales, where $\alpha$ is overestimated by DFA1 and 
MDFA1, see Figs. \ref{long}(b)-(d)). The upper limit in this "fixed lower limit" range is 
set to $N/2$ here.  Figure \ref{short} shows the result for this first definition. 
As can be clearly seen, the exponents become more accurate if a larger scaling range is 
used in the fitting procedure. While CMA and DFA1 show similar results and systematically 
underestimate the real scaling exponent for very short data ($N<100$)\footnote{This outcome
is rather surprising, since from Fig. \ref{long} one would expect that DFA1 and MDFA1 overestimate
$\alpha$ for short data (due to the deviations on small scales). However, it turns out
that one has to take into account the different averaging procedures. In Fig. 2 we simply average over 
all calculated exponents for given data lengths, which have a large standard deviation for short series. 
On the other side, in Fig. 1, the fluctuation functions are averaged non-logarithmically. 
Configurations with large values thus affect the means much more than 
configurations with small values. This averaging procedure favours larger slopes, 
since the variations of $F(s)$ are generally larger for larger $s$. Further on, the 
larger slopes occur for $s<10$ in Fig. 1(b), while the fitting range is set to 
$10\leq s\leq 25$ for the first point in Fig. 2(a).}, $\bar{\alpha}_{\rm 
MDFA}$ is quite stable. However, $\sigma(\alpha_{\rm MDFA})$ is significantly increased 
for $N<100$ (see also Fig. \ref{SD single}).

An alternative definition of $\alpha$ is based on a moving fitting regime with "fixed 
width", e.g., from $N/20$ to $N/2$. In this case, $\sigma(\alpha)$ is practically 
independent of $N$ (not shown). Since both definitions are identical for $N=200$, the
results in Fig. \ref{short} for $N=200$ are valid also for larger $N$ in case of the fixed
width definition. In the following, we will only refer to a fitting range with fixed lower limit.

An interesting question when studying the behaviour of $\sigma(\alpha)$ versus $N$ is
whether the variations of $\alpha$ are due to fluctuating properties of the data or due
to the inaccuracy of the methods. It is hard to clarify this, since there is no way to 
identify or define a scaling exponent for any data without applying an analyzing method.
Here we use DFA as such reference method. In the Fourier Filtering Method (see above)
the data is generated by manipulating the slope in the power spectrum, i.e. 
$\beta$, which is directly related to the exponent $\alpha$ (see Section 
\ref{sec crossover} and \cite{shlomo88}). Nevertheless, the shorter a time series, 
the less well-defined its intrinsic exponents $\alpha$ and $\beta$ become. 
%This is partly similar to wave mechanics -- limit a wavepacket to small space (time) 
%and its width in momentum (frequency) space becomes broad. 
The power spectrum and DFA fluctuation function become less smooth as a time series 
becomes shorter, increasing the error in calculating (and already defining) the 
exponents. The less smooth the curves are, the less accurate are the exponents defined %If the plot is not smooth, 
%the exponent is not exactly defined 
and different methods will yield different results.  This is essentially not an error 
in one or the other method.

\begin{figure} [ht] \begin{center}
\includegraphics[height=4.75cm]{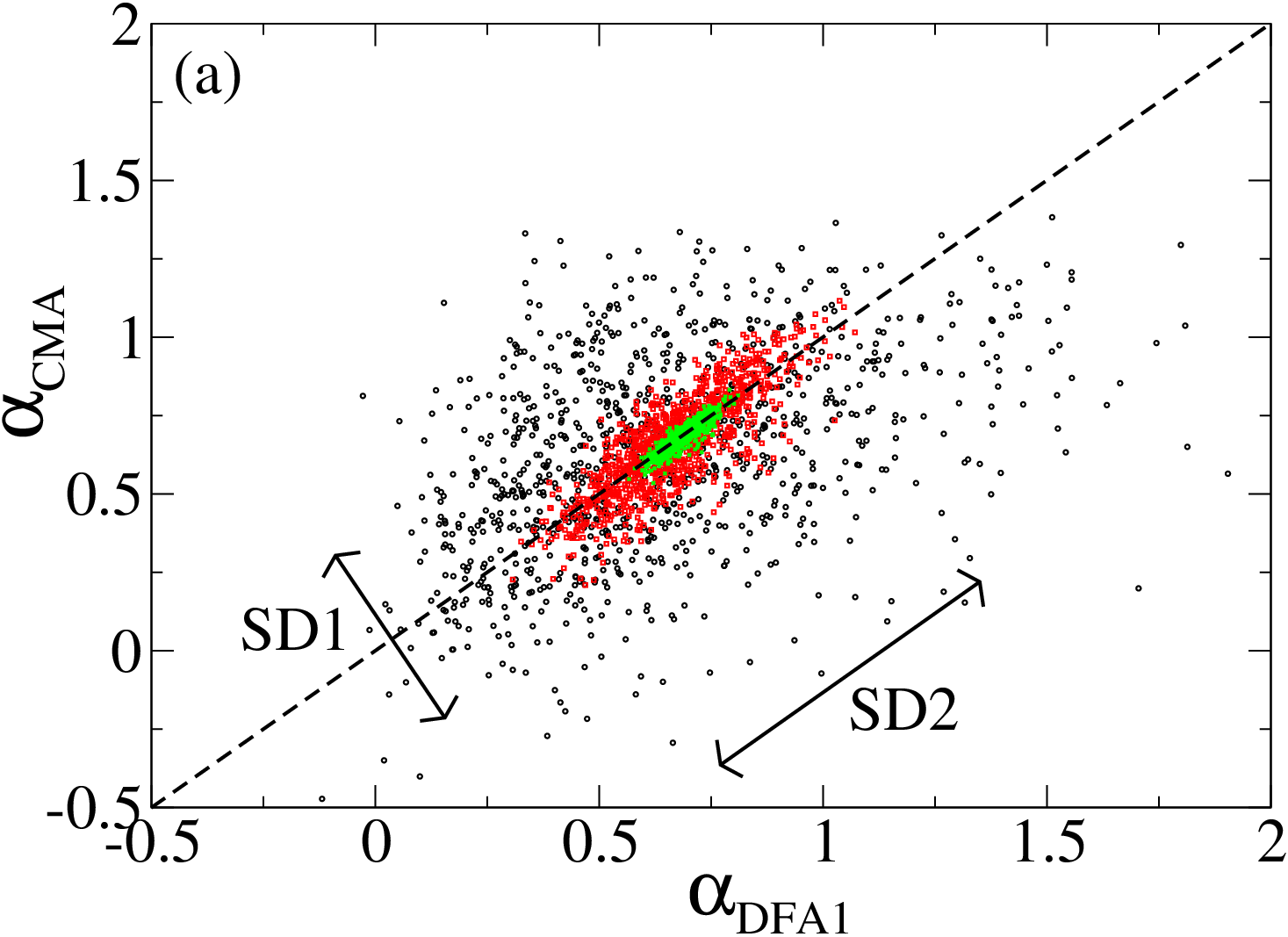}\hfill
\includegraphics[height=4.75cm]{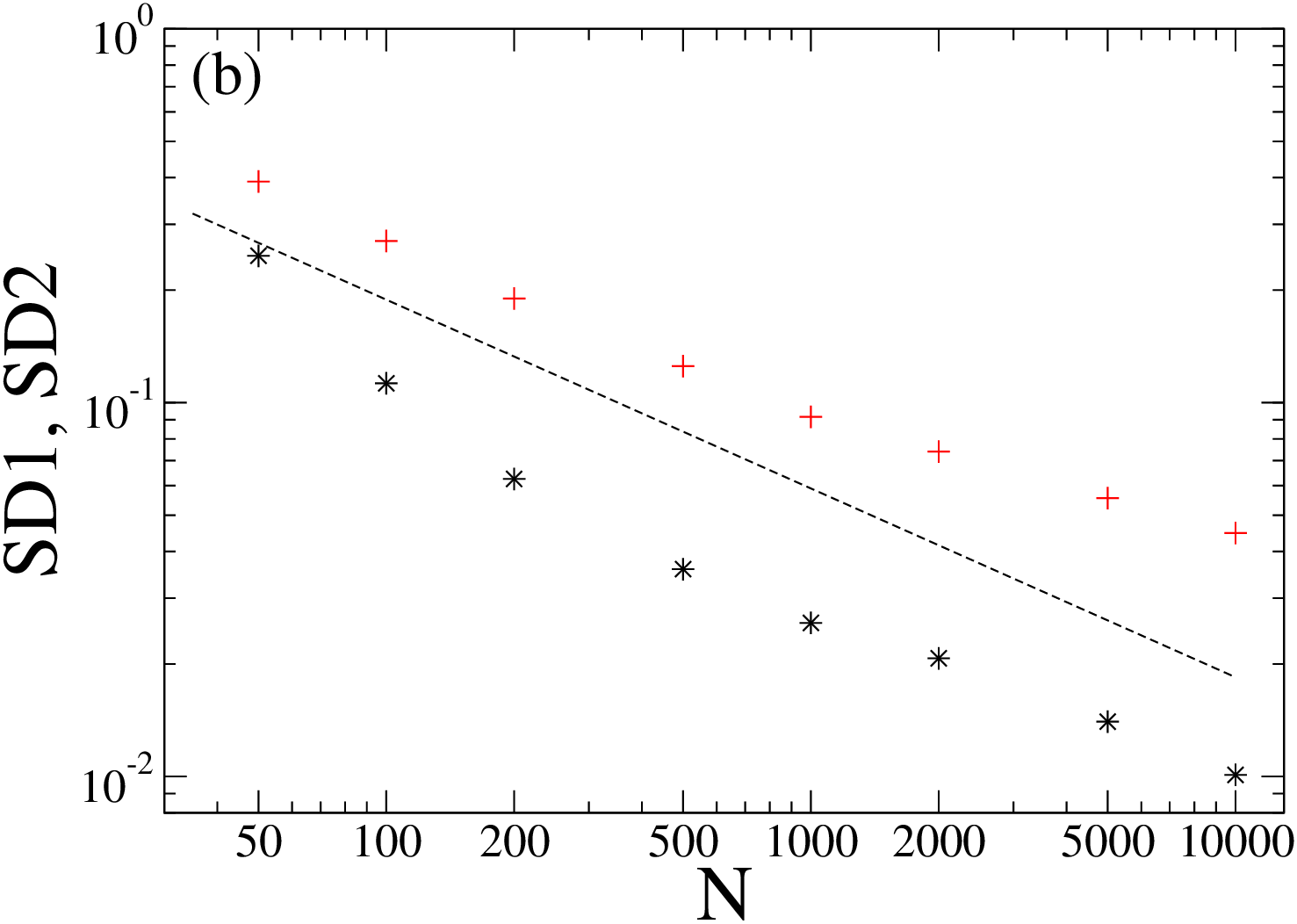}
\caption{(colour online) (a) Correlation scattering plots of the scaling exponents calculated with DFA1 
and CMA for $N=50$ (black), $N=200$ (red) and $N=5000$ (green). 
(b) Corresponding standard deviations $SD1$ (black stars) and $SD2$ (red plus signs) as defined in Eqs.~(\ref{SD}) 
versus $N$. Note that $SD1$ decreases faster with $N$ than $SD2$, 
suggesting that the uncertainty of CMA and DFA1 decreases faster with 
$N$ than the indeterminacy of data generation; for comparison see the dotted line: 
$SD \sim (\ln N)^{-1/2}$. The results of 1000 series with an imposed value 
of $\alpha=0.7$ are shown.}
\label{poincare DFA-CMA} \end{center} \end{figure}

\begin{figure} [t] \begin{center}
\includegraphics[height=4.75cm]{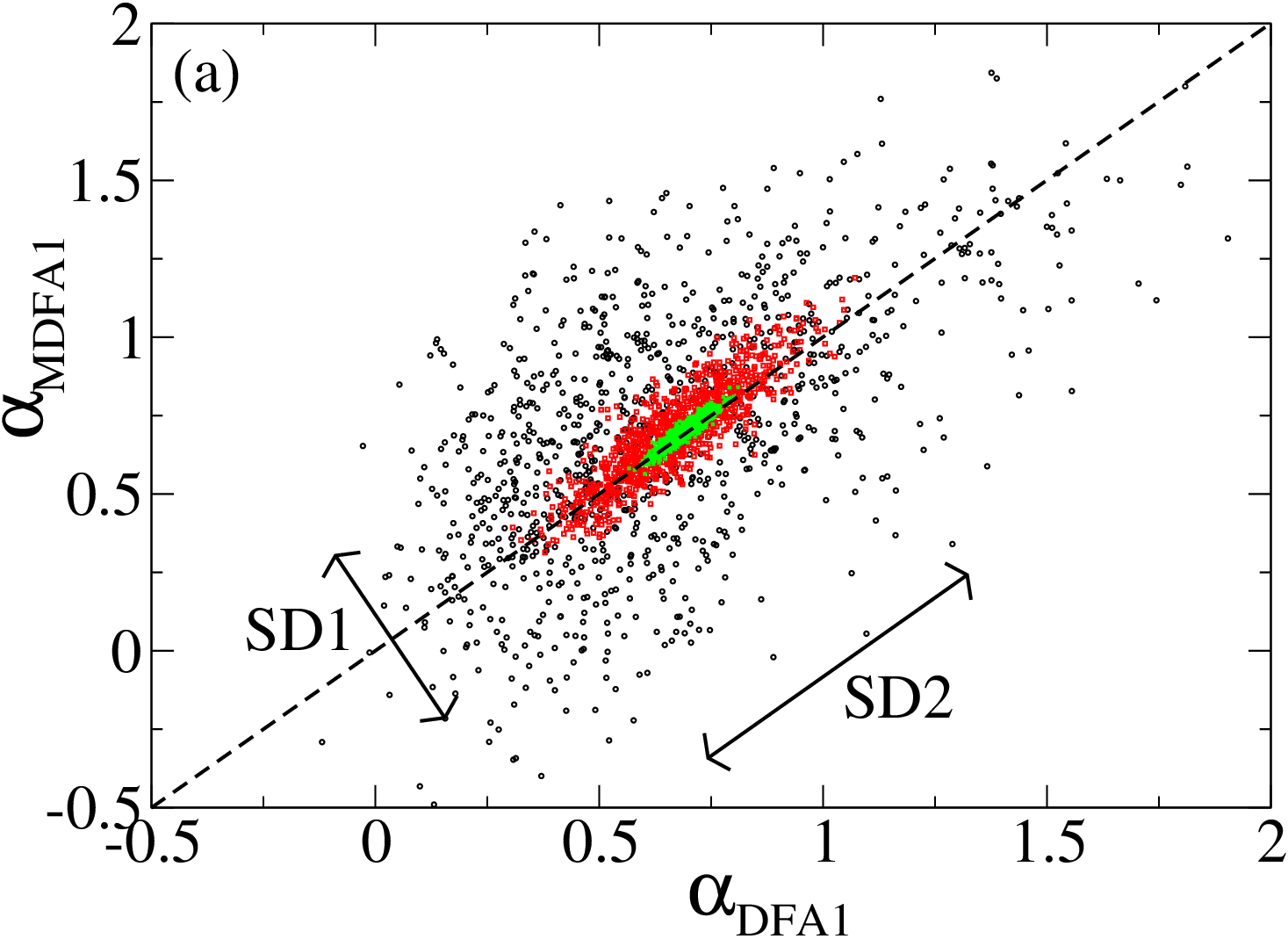}\hfill
\includegraphics[height=4.75cm]{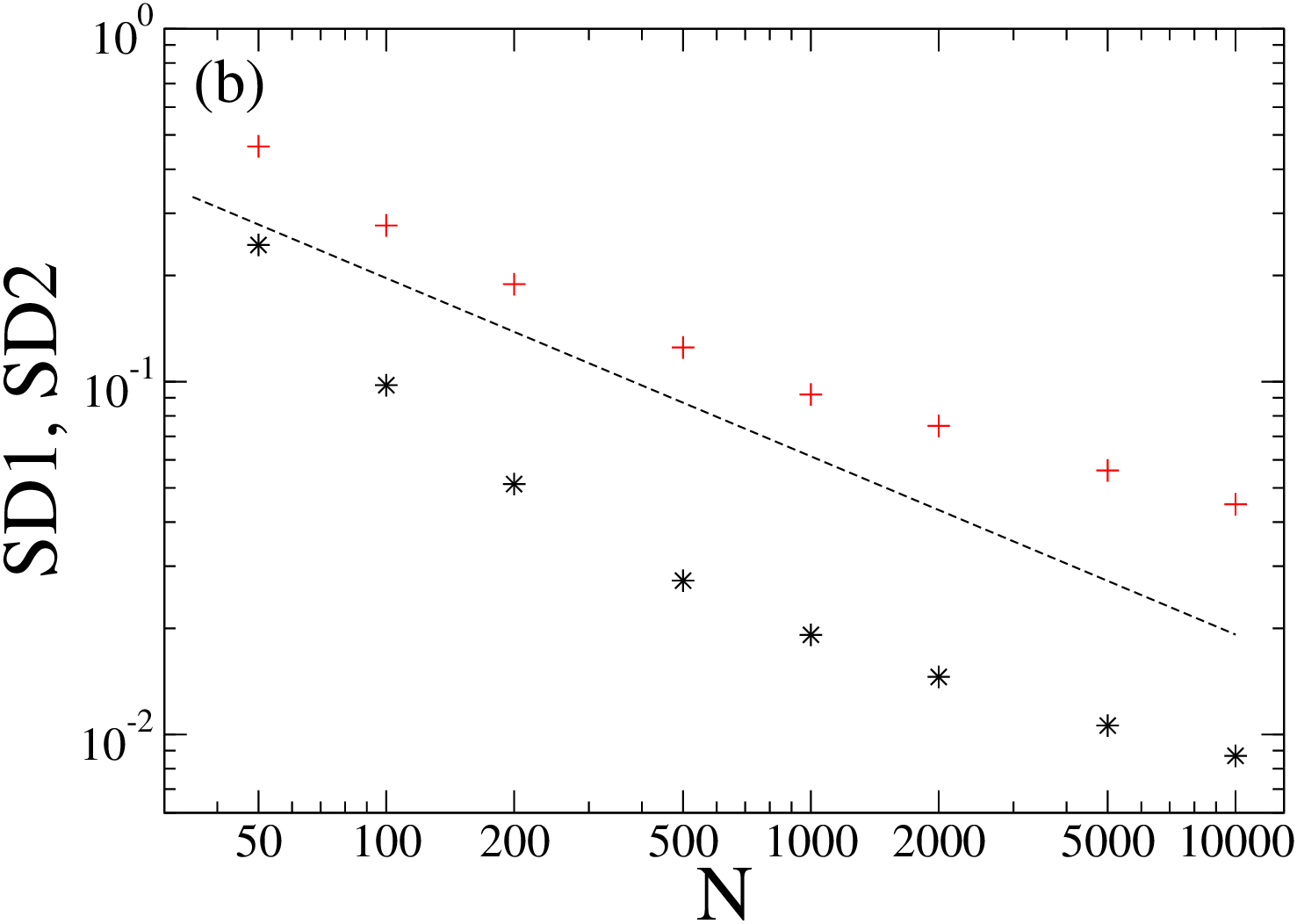}
\caption{(colour online) Same as Fig. 4 but for the scaling exponents calculated with DFA1 and MDFA1.
Similar plots are obtained for CMA versus MDFA1 (not shown).}
%$N=50$ (black circles), $N=200$ (red squares) and $N=5000$ (green crosses) and (b) 
%corresponding standard deviations
%$SD1$ and $SD2$ (see Eq.~(\ref{SD})) versus $\ln N$. Also in this case SD1 (black stars) 
%decreases much faster with the signal length than SD2 (red plus signs); 
%for comparison see the dashed line: $SD\propto (\ln N)^{-3.0}$.)}
\label{poincare DFA-MDFA} \end{center} \end{figure}

In order to get an impression of the uncertainty of the methods and the error
in generating data sets with a certain scaling exponent, one can study correlation
scattering plots (Figs. \ref{poincare DFA-CMA} and \ref{poincare DFA-MDFA}). The
standard deviations to characterize such plots are defined by \cite{poincare}
\begin{equation}\begin{array}{ll}
SD1 &= \sqrt{\frac{1}{N}\sum_{i=1}^N\frac{1}{2}[(\alpha^i_{\rm DFA}-\alpha^i_{y})-
\erw{\alpha^i_{\rm DFA}-\alpha^i_{y}}]^2},\\
SD2 &= \sqrt{\frac{1}{N}\sum_{i=1}^N\frac{1}{2}[(\alpha^i_{\rm DFA}+\alpha^i_{y})-
\erw{\alpha^i_{\rm DFA}+\alpha^i_{y}}]^2},
\end{array}\label{SD}\end{equation}
so that $SD1$ ($SD2$) is the standard deviation perpendicular (parallel) to the line given 
by $\alpha_{\rm DFA}=\alpha_{y}$, where $\alpha_y$ is the scaling exponent calculated by method 
$y$. Assuming that $\alpha_y$ is composed of the intrinsic scaling exponent 
$\tilde{\alpha}_y$ of method $y$ and an error $\Delta\alpha$ because of data generation, 
it can be seen from Eq.~(\ref{SD}) that $SD1$ eliminates $\Delta\alpha$ and thus may give 
a hint about the accuracy of method $y$.\footnote{Alternatively, if one does
not compare method $y$ to a reference method, such as DFA1 in this case, 
an additional error by estimating $\alpha$ should
be taken into account.} If the considered method calculates exactly 
the same scaling exponent as DFA1, $SD1$ would vanish. On the other hand, if method $y$
deviates from DFA1, $SD1$ will become large, i.~e. comparable with $SD2$.  Consequently, 
the indeterminacy of data generation can be assessed by the difference between $SD2$ and $SD1$. 

Figures \ref{poincare DFA-CMA} and \ref{poincare DFA-MDFA} show that $SD1$ is clearly smaller 
than $SD2$ suggesting that the errors from data generation are larger than the deviations of 
both, CMA and MDFA1 results from DFA1 results. In addition the decay of $SD1$ is faster than
$(\ln N)^{-1/2}$ while the decay of $SD2$ is slower than this. 

\begin{figure} [ht]\begin{center}
\includegraphics[height=4.9cm]{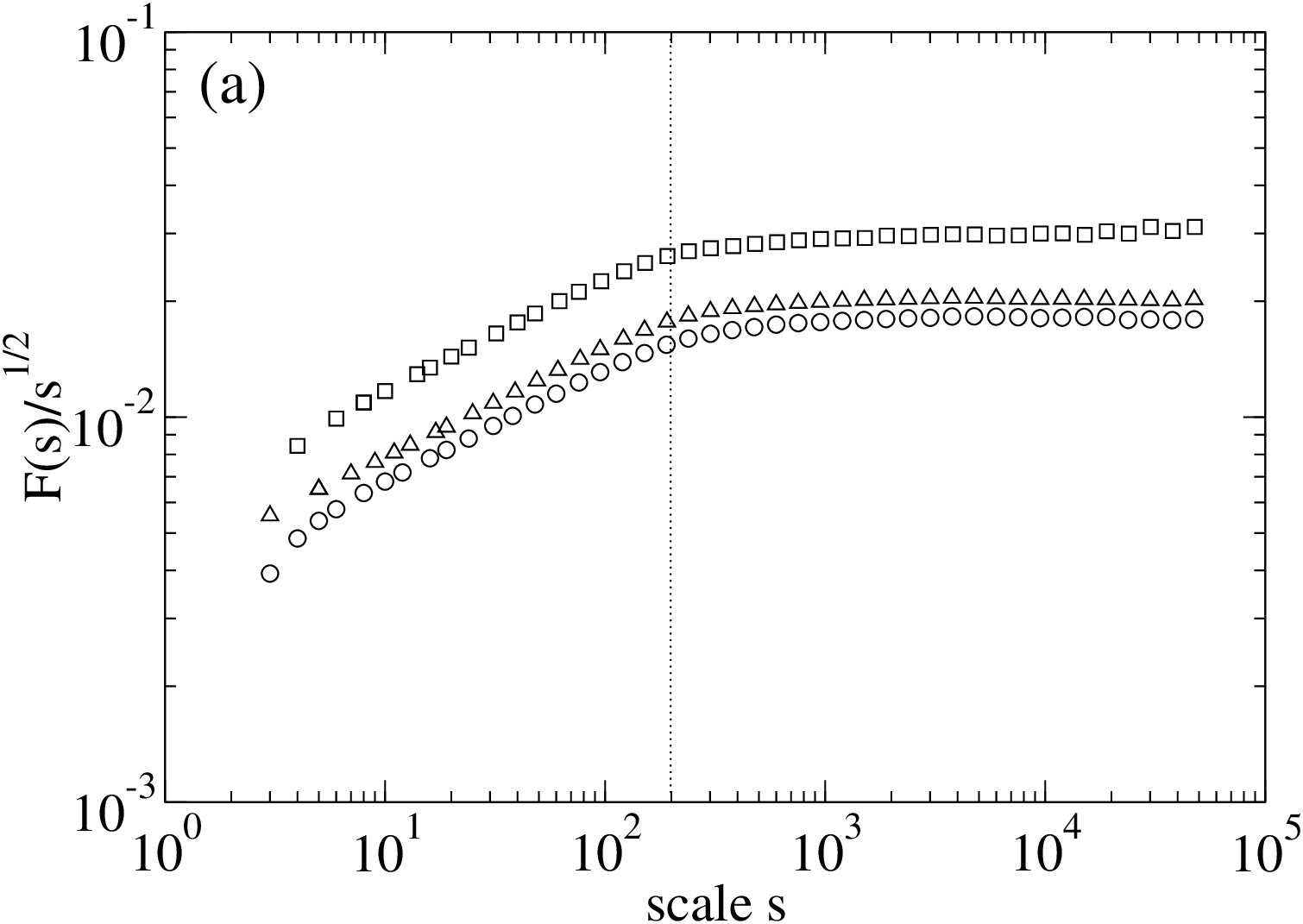}\hfill\includegraphics[height=4.8cm]{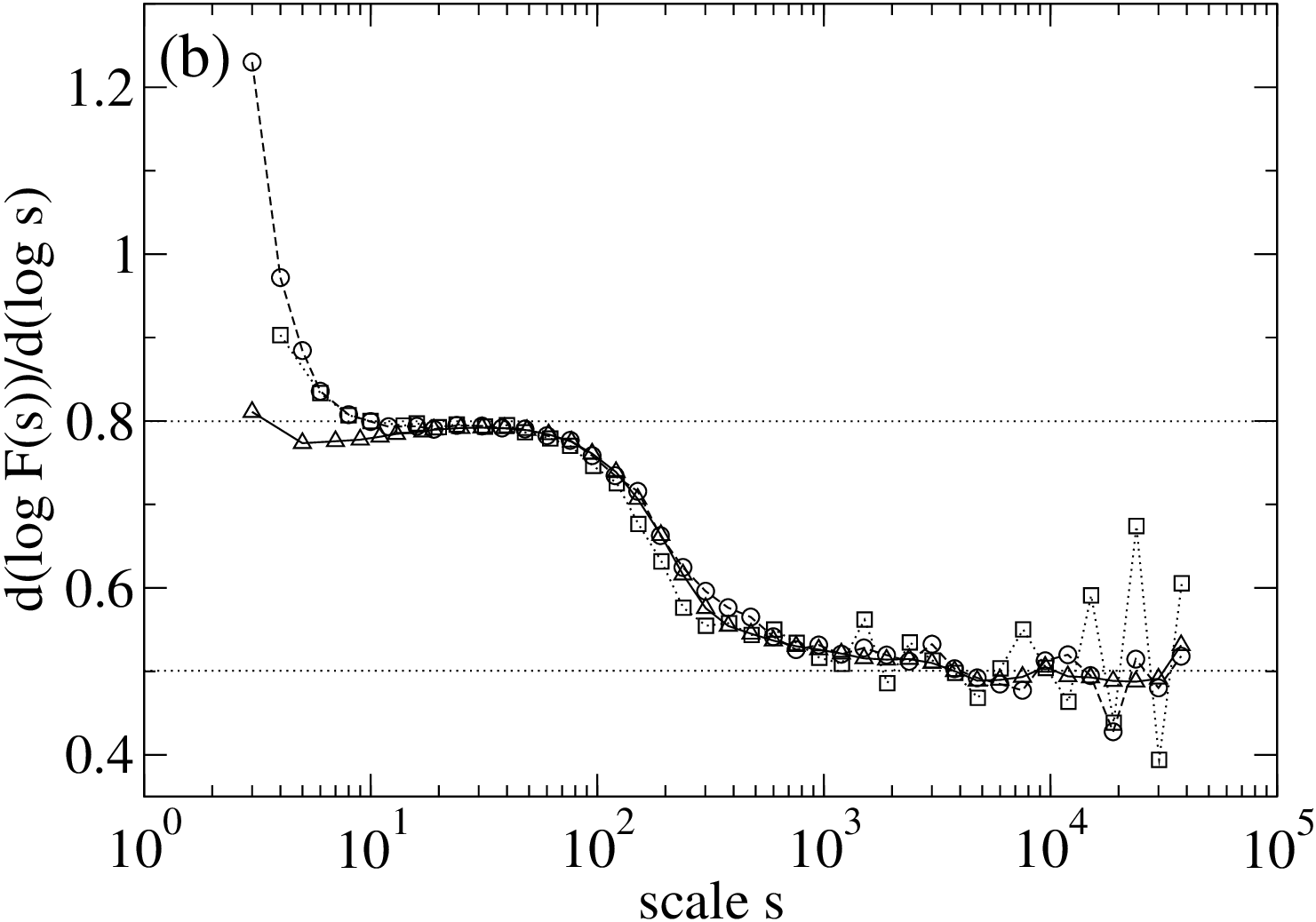}
\caption{(a) Fluctuation functions $F(s)$ of DFA1 (circles), CMA (triangles) and 
MDFA1 (squares) versus time scales $s$ for data with $\alpha=0.8$ for $s<s_\times$
and $\alpha=0.5$ for $s>s_\times$ (here $s_\times=200$). The results have been obtained 
by averaging 200 data series of length $N=100000$ for each method. Note that for the sake
of clarity $F(s)$ was divided by $s^{1/2}$. (b) Point to point 
derivative of $F(s)$ shown in (a).}
\label{crossover}\end{center}\end{figure}

\subsection{Determination of crossovers}
\label{sec crossover}

An often observed phenomenon in real world data sets is the occurrence of crossovers, 
i.e., the correlations of the recorded data do not follow the same scaling law for all 
time scales $s$. Crossovers occur, for example, in the analysis of short-term correlated 
data with finite decay time. Hence, an exact detection of crossovers is 
essential for finding characteristic time 
scales in complex systems. To compare the performance of DFA1, CMA, and MDFA1 regarding 
the detection of crossovers, we applied these methods to artificial time series with a 
well-defined crossover at scale $s_\times$. The results are shown in Fig. \ref{crossover}. 

It is sufficient to study only one scenario of a crossover in artificial data since the 
systematic deviation of the observed crossover from the real crossover was found to be 
independent of the values of $\alpha$ for DFA \cite{kantelhardt01}. A convenient way to generate 
such time series is by using a modification of the Fourier filtering method. If we need 
a crossover at scale $s_\times$ with scaling exponents $\alpha_1$ for $s<s_\times$ and 
$\alpha_2$ for $s>s_x$, the power spectrum of an uncorrelated random series is multiplied 
by $(f/f_\times)^{-\beta_2}$ for low frequencies $f<f_\times=1/s_\times$ and with 
$(f/f_\times)^{-\beta_1}$ for frequencies $f>f_\times$.  The relation between $\alpha$ 
and $\beta$ is given by $\beta=2\alpha-1$ \cite{shlomo88}; the inverse Fourier transform 
of the manipulated power spectrum yields the desired data.

\begin{figure} [ht] \begin{center}
\includegraphics[height=6cm,angle=0]{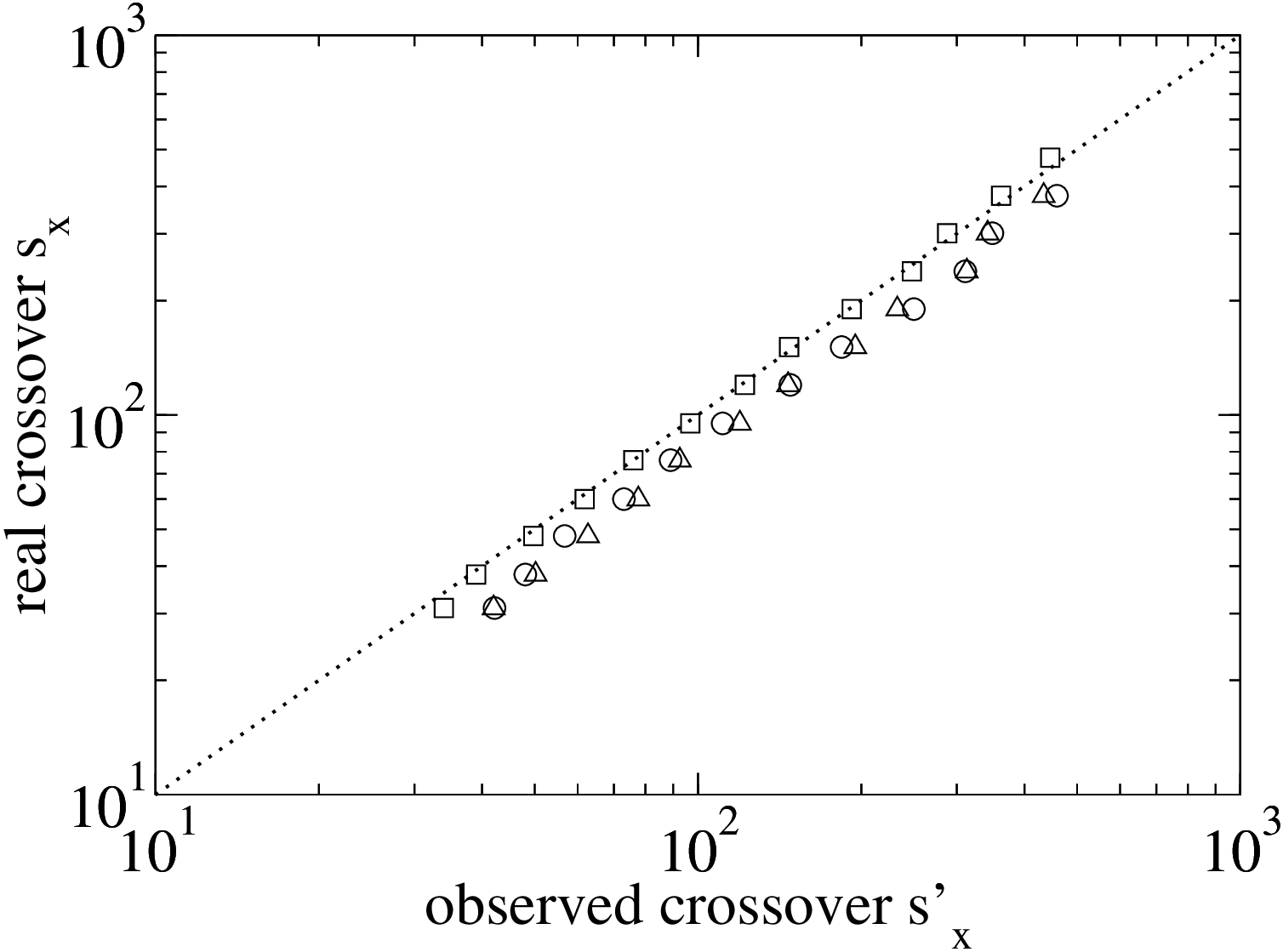}
\caption{Real crossovers $s_\times$ versus observed crossovers $s'_\times$ for DFA1 
(circles), CMA (triangles) and MDFA1 (squares). The results have been obtained by
averaging over the same number of configurations as in Fig. 6 for each $s_\times$. 
The $s_\times$ values can be estimated from $s'_\times$ by 
$\ln s_\times\approx\ln s'_\times -0.25$ (DFA1), $\ln s_\times\approx 1.05\ln s'_\times -0.47$ (CMA), 
and $\ln s_\times\approx 1.04\ln s'_\times -0.19$ (MDFA1), respectively.}
\label{crossover2} \end{center} \end{figure}

Figure \ref{crossover2} shows our results for generated data with systematically varied 
real crossover $s_\times$. While DFA1 and CMA slightly overestimate the position of the 
crossover ($s'_\times > s_\times$) by the same degree, MDFA detects $s_\times$ rather 
accurately. Clearly, a linear relationship between $s_\times$ and $s'_\times$ is observed.
Hence, when observing a crossover at position $s'_\times$ in DFA1, CMA, or MDFA1, the real
crossover position $s_\times$ can be estimated by the equations given in the caption
of Fig. \ref{crossover2}.   

\subsection{Data with monotonous trends}

Trends are ubiquitous in many noisy signals obtained from real
systems. As it was discussed above and shown in previous work, trends may mask 
the real correlation behaviour of the intrinsic fluctuations in the data. To
study the effect of trends in DFA, CMA and MDFA, we have added a linear 
and a non-integer trend to the original record $(x_i)$ generated with the 
Fourier transform method. Other kinds of trends (polynomial, sinusoidal and 
irregularly oscillating) have been systematically studied elsewhere 
\cite{kantelhardt01,hu01,EMD-DFA,SVD-DFA1,SVD-DFA2}.

\begin{figure} [ht] \vspace{0.5cm} \begin{center}
\includegraphics[height=5.0cm]{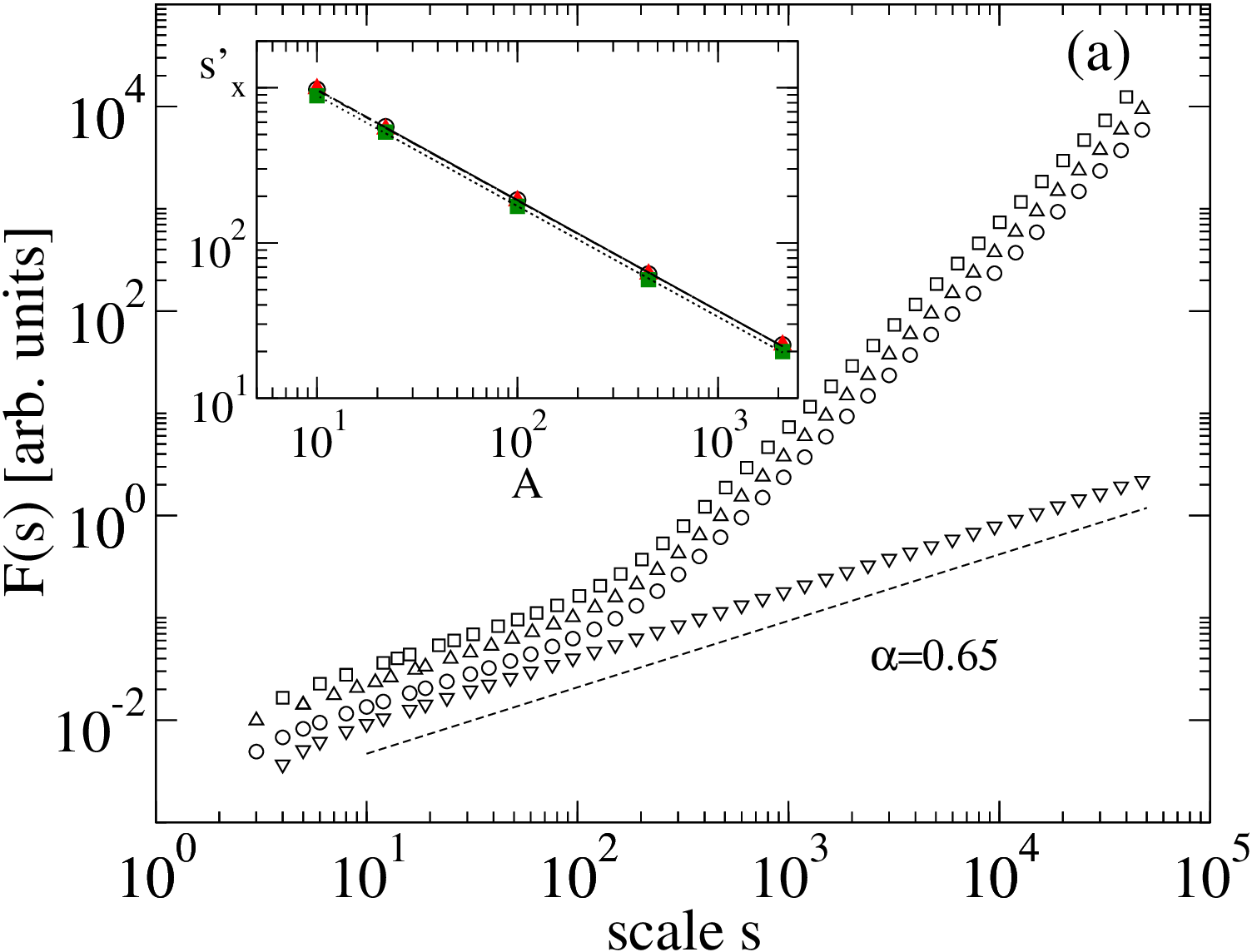}\hfill\includegraphics[height=5.0cm]{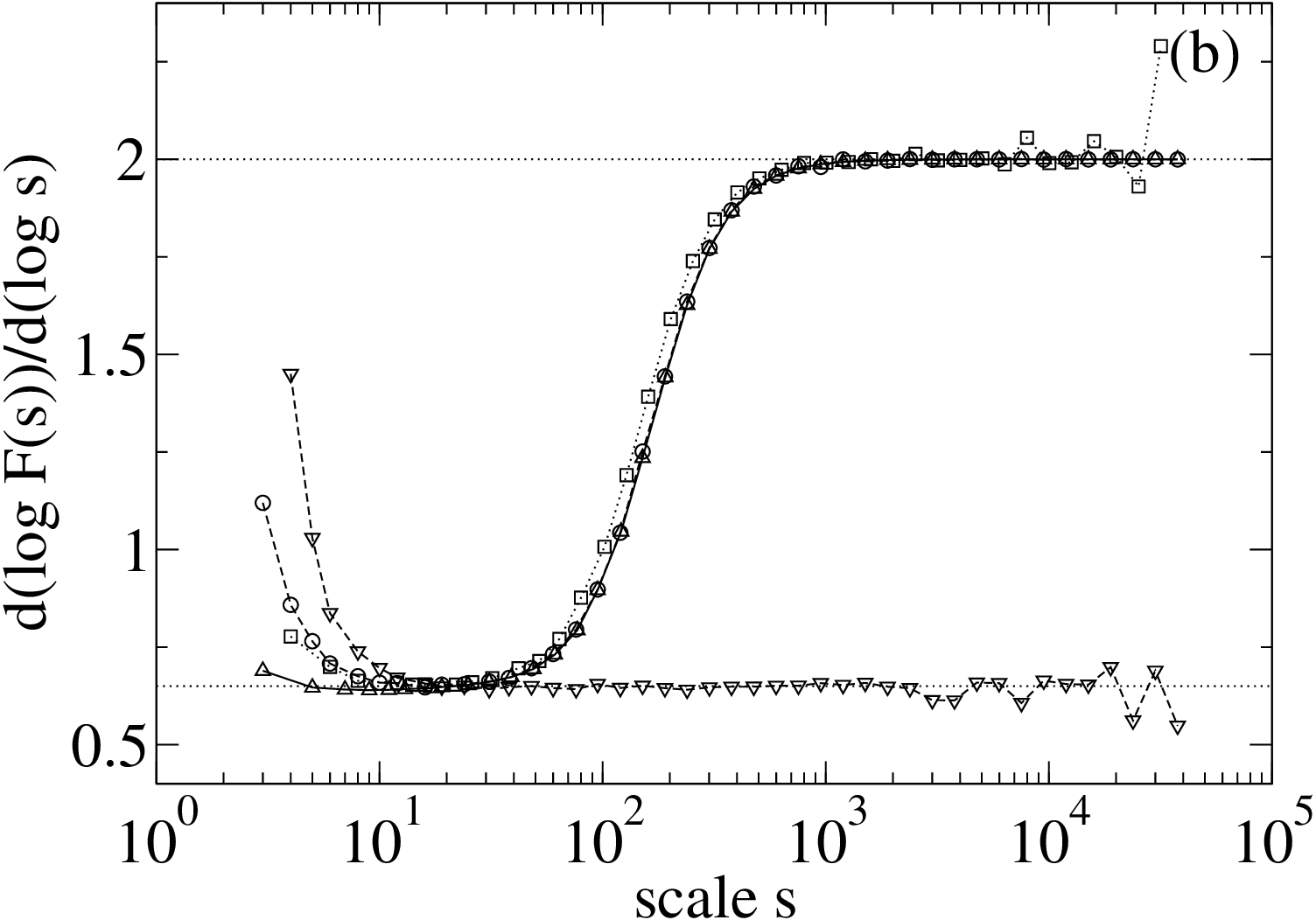}
\caption{Fluctuation functions and point to point derivative of long-range correlated data sets
with a linear trend: $x^\prime_i=x_i+Ax$ with $x=i/N$ ($\alpha=0.65$, $A=10$). 
(a) Trend-related crossover after analysis by 
means of DFA1 (circles, $s'_\times\approx 187$), CMA (triangles up, $s'_\times\approx 186$) and MDFA1 
(squares, $s'_\times\approx 170$). For comparison, $F(s)$ versus $s$ is shown also for DFA2 (triangles 
down). Since DFA2 eliminates the linear trend in $x^\prime_i$ we find the expected scaling 
behaviour $F(s)\sim s^{0.65}$ without crossover (the DFA curves were shifted 
downwards for better visibility). In the inset the position of the crossover is shown for 
different strengths $A$ of the trend for all three methods; the data can be fitted by 
$s'_\times\propto A^{-\delta}$ with an exponent $\delta \approx 0.71$, independent of the
technique. (b) Point to point derivative of the $F(s)$ functions 
shown in (a). The results were obtained by averaging 100 data series of 
length $N=100000$.}
\label{trend}\end{center}\end{figure}

Figure \ref{trend} depicts our results after adding a linear trend to long-range
correlated data (with $\erw{x}=0$ and $\sigma(x)=1$ before adding the trend). 
Since DFA1, CMA and MDFA1 are, by definition, not able to remove 
linear trends in the original data, all $F(s)$ curves show trend induced crossovers 
at $s'_\times$, which occur slightly earlier for MDFA1.  
Above the crossover, an artificial scaling exponent 
$\alpha_{\rm trend}=2$ is observed in agreement with \cite{kantelhardt01}.  
A systematic variation of the strength of the trend $A$ shows that the crossover 
position $s'_\times$ increases with $A$ as $s'_\times \sim A^{-\delta}$ with an 
exponent $\delta \approx 0.71$, independent of the technique (see inset 
of Fig. \ref{trend}(a)) and also independent of the fluctuation exponent $\alpha$ 
(not shown). This scaling relation allows to extrapolate for smaller values of $A$.
For comparison, the fluctuation function $F(s)$ is also shown for 
DFA2. In this case, we clearly observe the expected scaling behaviour without any
crossover.

The application of a non-integer trend, e.~g. $x^\prime_i=x_i+A(i/N)^{1.2}$
leads to similar results as shown in Fig. \ref{trend} (not shown, see also \cite{kantelhardt01} for DFA).  
However, here we also observe a trend related crossover for DFA2.  

\section{Conclusion}
In summary, we have compared several recently suggested detrending methods
based on random walk theory which were 
developed to detect long-range correlations in data affected by trends. In particular, we 
investigated the performance of the Centered Moving Average (CMA) method and the 
Modified Detrended Fluctuation Analysis (MDFA) regarding their behaviour on small 
and large scales, and the determination of crossovers in monofractal data sets with different 
lengths and monotonous trends. A systematic comparison of CMA and MDFA with standard 
DFA showed a small advantage of CMA in the computation of the scaling behaviour on 
small ($s<10$) and large ($s>N/4$) scales.  The detection of crossovers in the data 
was somewhat more exact with MDFA.  Ultimately, we think that CMA is a good alternative 
to DFA1 when analyzing the scaling properties in short data sets without trends. 
Nevertheless for data with possible unknown trends we recommend the application of 
standard DFA with several different detrending polynomial orders to distinguish 
real crossovers from artificial crossovers due to trends. In addition, an independent
approach (e.g., wavelet analysis) should be used to confirm findings of long-range
correlations.

{\it Acknowledgement:}
We thank Diego Rybski for very helpful discussions. 
This work has been supported by 
the Deutsche Forschungsgemeinschaft (grant KA 1676/3) and the European Union 
(STREP project DAPHNet, grant 018474-2). RB acknowledges financial support 
from the President scholarship of Bar-Ilan University.

\end{document}